\documentclass[11pt,3p,twocolumn,authoryear]{elsarticle}

\journal{Global and Planetary Change}

\usepackage[utf8]{inputenc}		%
\usepackage[T1]{fontenc}		%
\usepackage{graphicx}			%
\usepackage{amssymb} 			% Symbols
\usepackage{amsmath} 			% Symbols
\usepackage{wasysym} 

\usepackage{lineno}
\begin{document}

\begin{frontmatter}

\title{A physical framework for the Earth System, Anthropocene Equation and the Great Acceleration}

\author{O. Bertolami}
\ead{orfeu.bertolami@fc.up.pt}

\author{F. Francisco}
\ead{frederico.francisco@fc.up.pt}

\address{Departamento de Física e Astronomia and Centro de Física do Porto, Faculdade de Ciências, Universidade do Porto, Rua do Campo Alegre 687, 4169-007 Porto, Portugal}

\begin{abstract}
	It is proposed, based on the Landau-Ginzburg Theory of phase transitions, that the transition of the Earth System from the stable conditions of the Holocene to the human driven condition of the Anthropocene is, actually, a phase transition, a qualitative change away from its Holocene equilibrium state. Based on this physical framework, one obtains the Anthropocene equation, the so-called Great Acceleration and shows that (i) the Earth System temperature on the new equilibrium state diverges from the average temperature of the Holocene as the cubic root of the human intervention, described by a parameter, $H$; (ii) the human induced departure from the Holocene can be as drastic as the ones due to  natural, astronomical and geophysical causes; (iii) the susceptibility of the Earth System to human effects is much more relevant near the phase transition. The procedure to obtain numerical predictions from data is also exemplified through one of the existing proposals to account for human impact on the Earth's Holocene equilibrium.
\end{abstract}


\end{frontmatter}

%-----------------------------------------------------------------------------%
%-- Body ---------------------------------------------------------------------%
%-----------------------------------------------------------------------------%

Declarations of interest: none

\section{Introduction}

The Earth System (ES) is an encompassing planetary system that comprises the biosphere, \textit{i.e.}, the sum of all living biota and their interactions and feedbacks with the geosphere, including the atmosphere, hydrosphere, cryosphere and upper lithosphere \citep{Steffen:2016}.

Over its evolution, the ES has been driven predominantly by astronomical and geophysical forces. However, in the last six decades or so, one has witnessed a large increase in the rate of growth of the anthropological impacts \citep{Steffen:2014}. This recent trend is further highlighted when put against the remarkable stability of last ten millennia \citep{Ganopolski:2001,Jones:2004}, the epoch that started with the end of the last glacial period, the Holocene. This has led to the hypothesis that the Earth has entered a new geological epoch, driven by the changes due to human activities, the Anthropocene \citep{Crutzen:2000,Crutzen:2002,Steffen:2007,Steffen:2011,Waters:2016}.

In this context, it has been recently proposed that the rate of change of the ES in the Anthropocene could be captured by an abstract equation, the Anthropocene Equation (AE), where the ES is depicted as a function, $E$, for which the rate of change is driven by natural and human forces: astronomical, $A$, geophysical, $G$, internal dynamical forces, $I$, and human activity effects, generically denoted by $H$. Human intervention is represented as a function of the population of consumers, $P$, consumption of resources, $C$, and the ``Technosphere''. The Techosphere, $T$, is further decomposed into the energy system, $E_n$, the degree of knowledge, $K$, and political economy, $Pe$ \citep{Gaffney:2017}.

In this paper, it is argued that the AE must arise from a thermodynamical description of the state of the ES in terms of the driving forces described above. These forces characterize the thermodynamic state of the system in terms of the Helmholtz free energy, $F(A,G,I,H)$. This is a natural choice since, from the free energy, one can obtain all relevant thermodynamic variables. The natural variables will be collectively denoted as $\eta = (A,G,I)$. The rate of change of $F(\eta,H)$ over time is then given by
\begin{equation}
	{dF \over dt} = {\partial F \over \partial t} + \left({\partial F \over \partial \eta} \right)_H {d\eta \over dt} + \left({\partial F \over \partial H} \right)_{\eta} {dH \over dt}.
\end{equation}

The AE arises from the unquestionable observation that, from the second half of the last century onward, changes to the ES have been driven predominantly by the human influences, that is, $({\partial F \over \partial H})_{\eta} \gg ({\partial F \over \partial \eta})_H$, and thereby, further assuming that the free energy has no explicit dependence on time, one obtains the AE
\begin{equation}
	{\setlength\arraycolsep{2pt}
	\begin{array}{r c l}
		\displaystyle {dF \over dt} &\simeq& \displaystyle \left({\partial F \over \partial H} \right)_{\eta} {dH \over dt} \\
		&:=& f(H), \quad \eta = \text{const.},
		\label{eq:anthropoceneequation}
	\end{array}}
\end{equation}
\noindent where $f(H)$ is a function of $H$ with fixed $\eta$ values.

It will be seen that, from the AE, one can derive the Great Acceleration due to the human forcing. A physical framework that describes the passage from the Holocene to the Anthropocene as a phase transition in the formalism of the Landau-Ginzburg theory (LGT) will be set up. Using the terminology of the LGT, the climatic stability of the Holocene era corresponds to a ``symmetric'' phase, one that has a single stable equilibrium, and the Anthropocene corresponds to a transition to an ``asymmetric'' phase, disturbed by the human activities. These are described as an ``external field'' that sets the ES on a trajectory away from the Holocene and keeps shifting the equilibria of the ``asymmetric'' phase as the disturbance increases.

In the next section, the features of the LGT are briefly discussed. In section \ref{sec:phase_transition} the LGT is used to describe the transition from the Holocene to the Anthropocene. Finally, in section \ref{sec:discussion} some estimates are performed and conclusions are drawn.

%-----------------------------------------------------------------------------%

\section{Landau-Ginzburg theory}

Phase transitions occur when the features of the equilibrium state of a physical system change qualitatively. These usually reflect changes in the microscopic structure of the system, but can be described in terms of macroscopic thermodynamical variables.

The LGT describes phase transitions in physical systems through their free energy, using an expansion in terms of an order parameter, $\psi$. This order parameter is a function of some state variable of the system, like temperature or magnetization. The phase transition occurs between a symmetric and an asymmetric phase. In this context, ``symmetric'' means that there are fewer possible macroscopic equilibrium states \citep{Landau:1937,Ginzburg:1960}. The LGT provides a useful framework to describe phase transitions in a wide range of systems in condensed matter physics, materials science, quantum field theory and cosmology. 

Let us consider a generic continuous phase transition at a critical value, $r_{\rm c}$, of some state variable, $r$, between a symmetric state at $r > r_{\rm c}$ and an asymmetric state at $r < r_{\rm c}$. In the LGT, the free energy of the system is given by an analytic function of the order parameter that can be written in the form
\begin{equation}
	{\setlength\arraycolsep{2pt}
	\begin{array}{r c l}
		F(T) &=& F_0 + a(q)\psi^2 + b(q)\psi^4 \\
		& + & c(q)|\nabla\psi|^2 + d(q) \nabla^2\psi + \ldots,
	\end{array}}
\end{equation}
where $F_0$ is a constant. The coefficients $a(q)$, $b(q)$, $c(q)$ and $d(q)$ are, themselves, analytic functions of the state variable $r$ in its reduced form $q = (r-r_{\rm c})/ r_{\rm c}$: 
\begin{equation}
	{\setlength\arraycolsep{2pt}
	\begin{array}{r c l}
		a(q) &=& a_0 q + a_1 q^2 + \ldots, \\
		b(q) &=& b_0 + b_1 q + \ldots, \\
		c(q) &=& c_0 + c_1 q + \ldots, \\
    	d(q) &=& d_0 + d_1 q + \ldots,
	\end{array}}
	\label{eq:GLcoeff}
\end{equation}
where $a_0$, $b_0$, $c_0$ and $d_0$ are positive constants. It is worth pointing out that, in most physical problems, a predictive description is possible with just a few constants.

By minimizing the free energy function, one can find the equilibrium states of the system. Of course, for spatially homogeneous systems, the position dependent terms are eliminated from the problem.

%-----------------------------------------------------------------------------%

\section{Holocene to Anthropocene phase transition}
\label{sec:phase_transition}

One can regard the Holocene to Anthropocene transition as a continuous phase transition where local fluctuations drive the transition and become correlated over all distances, driving the ES towards the Anthropocene phase. As such, the properties of the Anthropocene can be described by an effective theory through a phenomenological Hamiltonian or a Free Energy function. This description is built upon LGT, as discussed in the previous section

The first step in setting up the LGT approach to this transition is the identification of the order parameter, $\psi$, which is chosen to be the ES temperature, $T$, with respect to the Holocene baseline temperature, $T_{\rm H}$, \textit{i.e.}, $\psi := (T - T_{\rm H})/T_{\rm H}$. Thus, one can then write the free energy as
\begin{equation}
	{\setlength\arraycolsep{2pt}
	\begin{array}{r c l}
		F(q) &=& F_0 + a(q)\psi^2 + b(q)\psi^4 \\
		&+& c(q)|\nabla\psi|^2 + d(q) \nabla^2\psi + \ldots,
		\label{eq:FreeEnergy}
	\end{array}}
\end{equation}
where $q = (\eta - \eta_0)/ \eta_0$, $\eta_0$ is the average values of the $A$, $G$ and $I$ forces during the Holocene, $F_0 = F(\eta_0)$ is a constant and $a(q)$, $b(q)$, $c(q)$ and $d(q)$ are analytic functions of $q$, analogous to Eq.\,(\ref{eq:GLcoeff}),
\begin{equation}
	{\setlength\arraycolsep{2pt}
	\begin{array}{r c l}
		a(q) &=& a_0q + \ldots,\\
		b(q) &=& b_0 + \ldots, \\
		c(q) &=& c_0 + \ldots, \\ 
    	d(q) &=& d_0 + \ldots .
	\end{array}}
\end{equation}
The constants $a_0,\ldots,b_0,\ldots,c_0, \ldots,d_0\ldots$ are positive, have dimensions of energy and can, in principle, be fitted from data. In fact, one shall see that it will be possible to describe the transition to the Anthropocene without the need to decompose the coefficients $a(q), b(q), \ldots$ in terms of the constants $a_0,\ldots,b_0,\ldots$. In what follows, the spatial dependence of $\psi$ is dropped, but its implicit time dependence is kept. Notice that odd power terms can also be included in the free energy. In fact, a linear term will be used to describe the human influence; cubic terms in the order parameter are suitable, for instance, to model metastability \citep{Paramos:2003}. 

Considering Eq.\,(\ref{eq:FreeEnergy}), one can find minima of the free energy in terms of the order parameter, corresponding to the ES equilibrium states for the two distinct phases:
\begin{equation}
	{\setlength\arraycolsep{2pt}
	\begin{array}{l}
	\displaystyle {dF \over d \psi} = 0 \Rightarrow \\
	\left\{
		\begin{array}{l l}
			\langle\psi\rangle = 0, 	& q \geq 0 \\
			\langle\psi\rangle= \left(-{a(q) \over 2 b(q)} \right)^{1 \over 2}, 	& q < 0 \\
		\end{array}
	 \right. ,
	 \label{eq:minimawithoutH}
	\end{array}}
\end{equation}
where the equilibrium states are denoted by $\langle\psi\rangle$.

\begin{figure}
	\centering
	\includegraphics[width=\columnwidth]{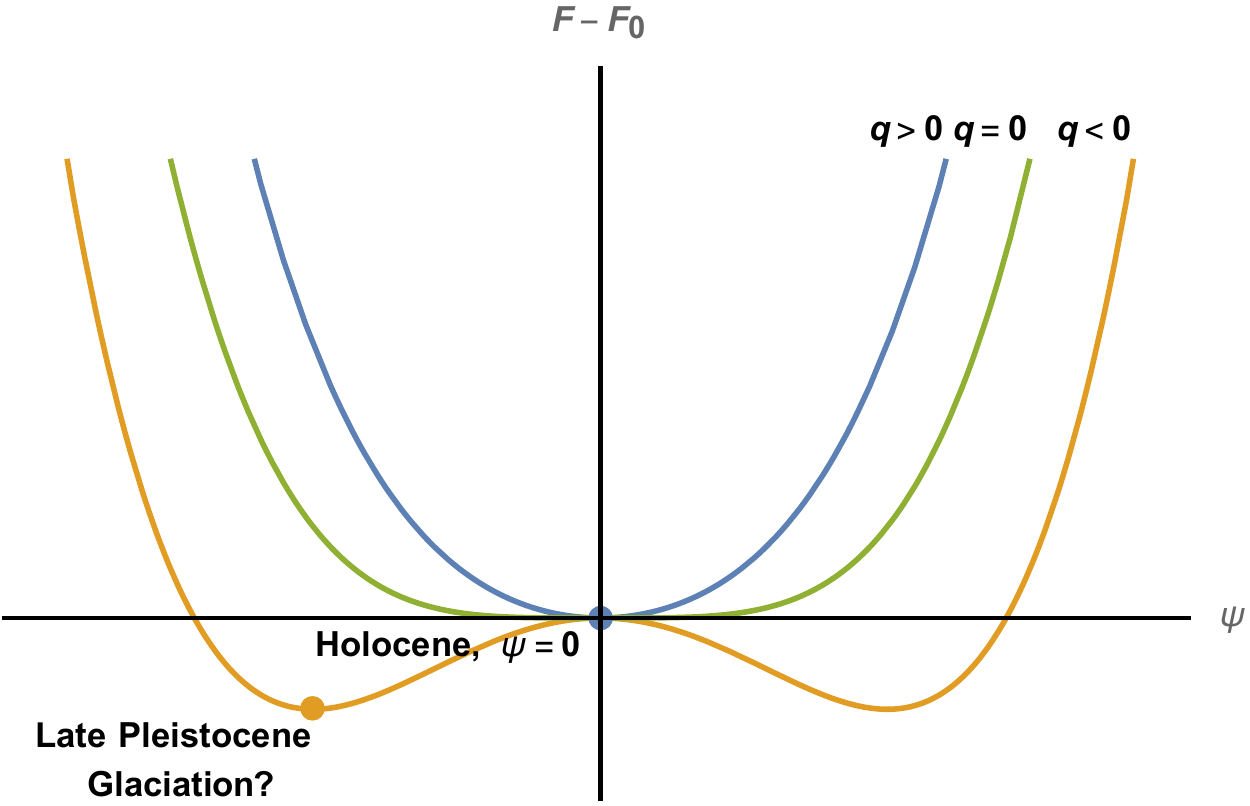}
	\caption{Free energy as a function of the order parameter for ES driven only by astronomical, geophysical and internal dynamics expressed in terms of the variable $q=(A,G,I)$. As all systems in a given phase tend to minimize their free energy, the Holocene stability is associated with the minimum of the ``symmetric'' phase $\langle\psi\rangle = 0$, for which $q>0$.}
	\label{fig:ESLGT}
\end{figure}

It is argued here that the stability of the Holocene is compatible with the ``symmetric'' phase $\langle\psi\rangle = 0$, with $q \geq 0$. As shown in Fig.\,\ref{fig:ESLGT}, the two phases are clearly distinguishable by their minima. The ``symmetric'' phase, with a single equilibrium state at $\langle\psi\rangle =0$, suggestively associated to the Holocene, with its characteristic stability. A change on the value of $q$ would trigger a transition to the ``asymmetric'' phase, where the ES would be led to evolve towards one of its minima, which could be interpreted as hotter interglacial periods or cooler glacial periods. This hypothesis actually fits quite well with the ``sawtooth'' dynamic of the ES during the Quaternary period, characterised by the the frequent oscillation between glacial and interglacial periods \citep{Lenton:2013}. In this model, this could correspond to an ES that is near the phase transition, with slight changes in $q$ knocking it slightly above or slightly below the threshold.

One accounts for the transition to the Anthropocene by considering an ``external field'', $H$, that models the effect of human activities and affects the free energy as follows:
\begin{equation}
	{\setlength\arraycolsep{2pt}
	\begin{array}{r c l}
		F(q,H) &=& F_0 + a(q)\psi^2 + b(q)\psi^4 \\
		&+& \ldots - h(q)H\psi ,
	\end{array}}
\end{equation}
where, like coefficients $a(q)$ and $b(q)$, $h(q)$ is an analytic function of $q$. This term is negative (assuming $h(q)>0$ and $H>0$) in order to match the observational fact that human intervention destabilises the ES towards warmer temperatures. The $h(q)$ coefficient can be absorbed by redefining $H$, so it will be dropped in the calculations that follow. 

The inclusion of human forcing changes the equilibrium equation 
\begin{equation}
	{\partial F(q,H) \over \partial \psi} = 0 \Rightarrow 2a\psi + 4b \psi^3 = H,
	\label{eq:freeeneregyminumum}
\end{equation}
leading to a different behaviour of the system near the phase transition. One can already forecast that human intervention can lead to effects of the same order of magnitude as natural ones if $|H| \simeq a(q)\psi$ or $|H| \simeq b(q)\psi^3$.

If one assumes the ES starts in the ``symmetric'' phase with $q \geq 0$, \textit{i.e.}, the Holocene, then Eq.\,(\ref{eq:freeeneregyminumum}) admits a single real solution given by
\begin{equation}
	{\setlength\arraycolsep{2pt}
	\begin{array}{r c l}
	\langle\psi\rangle &=& \left[ {H \over 8b} + \left({H^2 \over 64b^2} + {a^3 \over 216b^3} \right)^{1 \over 2} \right]^{1 \over 3} \\
	&+& \left[ {H \over 8b} - \left({H^2 \over 64b^2} + {a^3 \over 216b^3} \right)^{1 \over 2} \right]^{1 \over 3}.
	\label{eq:orderparamgenreal}
	\end{array}}
\end{equation}
In the case where the ES is close to the phase transition before human intervention, \textit{i.e.}, $q \simeq 0$ and $a(q) \simeq 0$, then this reduces to
\begin{equation}
	\langle\psi\rangle = \left( H \over 4b \right)^{1 \over 3}.
	\label{eq:orderparam}
\end{equation}
In both cases, the model predicts that the temperature of the new equilibrium departs from the the Holocene average, essentially, with the cubic root of $H$, the human intervention.

\begin{figure}
	\centering
	\includegraphics[width=\columnwidth]{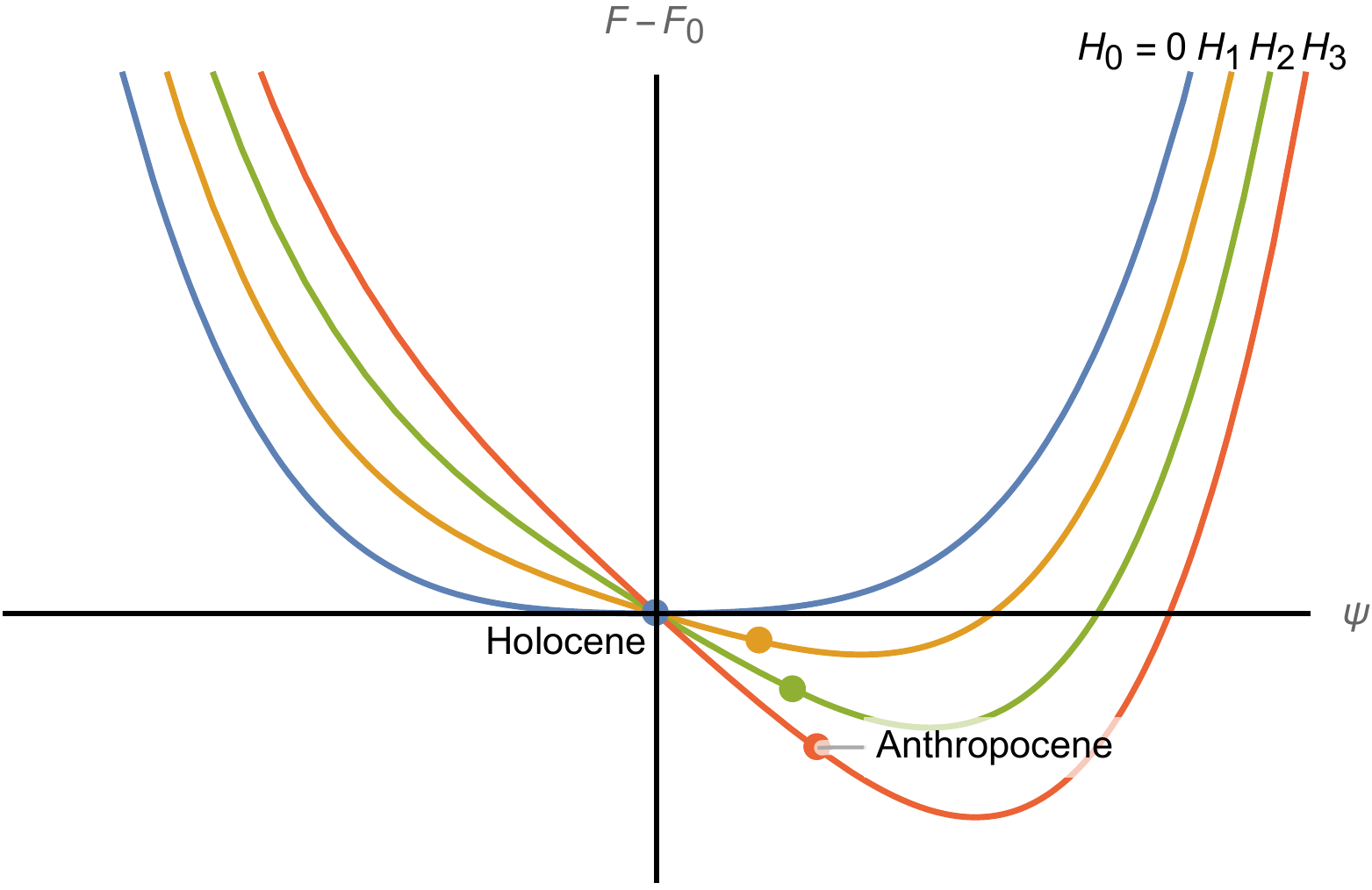}
	\caption{Free energy as a function of the order parameter for ES with increasing values of $H$, the human forcing external field. For $H_3>H_2>H_1>H_0=0$, the equilibrium point further departs from the Holocene stability, $\langle\psi\rangle = 0$, towards higher temperatures. The Anthropocene is a non-equilibrium state where the ES is moving towards a new yet undetermined equilibrium.}
	\label{fig:ESLGTH}
\end{figure}

This behaviour of the minimum is depicted in Fig.\,\ref{fig:ESLGTH}. The model clearly shows that the human forcing field is shifting the equilibrium point of the ES towards higher values of the order parameter, that is, towards higher temperatures. Furthermore, the free-energy well deepens as $H$ increases, and so does its slope, suggestive of the great acceleration that is indicated in the data \citep{Steffen:2014}. Indeed, from the AE, Eq. (\ref{eq:anthropoceneequation}), one can write for the acceleration of the change of the free energy
\begin{equation}
	\left|{d^2 F \over dt^2}\right| = {d \psi \over d t} {dH \over dt} + \psi {d^2 H \over dt^2}.
\end{equation}
This compares with the vanishing of $d^2 F/dt^2$ for $H=0$ at the symmetric phase, $\langle\psi\rangle = 0$. Hence, the LGT model of the ES is consistent with human forcing being the main driver of the changes in the ES.

To further understand the effect of the external field $H$ near the phase transition, one considers the susceptibility, a phenomenological quantity that can be computed in the LGT. In here, one defines the ES susceptibility to the Anthropocene, $\chi_{\rm A}$:
\begin{equation}
	\chi_{\rm A} \equiv \left({\partial \psi \over \partial H}\right)_{H \rightarrow 0}. 
\end{equation}
Then, from Eq.\,(\ref{eq:freeeneregyminumum}):
\begin{equation}
	\chi_{\rm A}^{-1} = 2 a + 12 b \psi^2 = \left\{
		\begin{array}{l l}
			2a, 	& q>0 \\
			-4a, 	& q<0 \\
		\end{array}
	 \right. .
\end{equation}
It is possible to verify that the susceptibility diverges close to the phase transition $\langle\psi\rangle = 0$. The consequence is the existence of a tipping point where the effects of the external field are much more important if the ES is close to the phase transition. On the other hand, according to the proposed model, if the ES was far from the phase transition, in either phase, the susceptibility would asymptotically vanish. In essence, the hypothesis that the ES changes due to natural or human causes can be verified through the divergence of $\chi_{\rm A}$.

Therefore, the model presented here already allows for three main predictions, namely: (i) if human forcing field keeps growing, the temperature will keep diverging from the Holocene one as $H^{1/3}$, as will its growth rate, since the equilibrium free energy shifts further below and away from the Holocene average; (ii) human intervention can lead to effects of the same order of magnitude as natural ones if $|H| \approx a(q)\psi$ or $|H| \approx b(q)\psi^3$; (iii) the susceptibility of the ES equilibrium to human intervention is much higher near the phase transition as $\chi_{\rm A}$ diverges, and conversely, it would tend to be negligible far from the phase transition.

%-----------------------------------------------------------------------------%

\section{Discussion and Outlook}
\label{sec:discussion}

The theoretical framework proposed here aims to provide a physical underpinning to the ES that can be tested with the available data. In this context, there are already some proposals on how to approach the modeling of the external field, $H$, the human forcing in the ES. It is worth stressing that the proposed model provides a quantitative and predictive framework for well founded qualitative ideas about the human domination of the biosphere.

One particular proposal is the one in which the human dominance of the ES is depicted as a net thermodynamic transformation of degrading the high density energy stored in the ``Earth-space battery'' into lower quality heat energy that is eventually radiated into deep space \citep{Schramski:2015}. It is argued that a specific fraction of this stored energy, the one in living biomass, is of critical importance for the sustainability of the biosphere, since it is the only one with a non-negligible replenishment rate through the flow of solar energy used by photosynthesis.

The predictability of our LGT based model can be verified when it is combined with a quantitative data based description of the natural and, particularly, the human drivers of the Earth system. In order to exemplify how this can be done, one uses the mentioned Earth-space battery paradigm from Ref.\,\citep{Schramski:2015}.

Taking into account only the living biomass stock, one can build the $H$ function as a measure of the depletion of the living biomass stock, in the simplest linear case:
\begin{equation}
	H = \alpha \Delta m_\text{LB} ,
	\label{eq:HBiomass}
\end{equation}
where $\Delta m_\text{LB}$ is a measure of the living biomass depletion, therefore a negative value, and $\alpha$ is an empirical coefficient that will account for the energy density of the living biomass, roughly $35 \times 10^6~{\rm J}$ per $1~{\rm kg}$ of carbon, and the effect of its change on the forcing of the ES.

The depletion rate of living biomass stocks has been much steeper in the last two centuries. It has decreased from an estimated $750 \times 10^{12}~{\rm kg}$ in 1800 to $660 \times 10^{12}~{\rm kg}$ in 1900 and $550 \times 10^{12}~{\rm kg}$ in 2000, showing an accelerated decline in the 20th century \citep{Smil:2011,BarOn:2018}. These figures can be used to estimate $\Delta m_\text{LB}$ in Eq.\,(\ref{eq:HBiomass}). This can then be used with Eqs.\,(\ref{eq:orderparamgenreal}) or (\ref{eq:orderparam}) to fit the coefficients $a(q)$ and $b(q)$ to the temperature record.

If one assumes the ES is close to the phase transition during the Holocene, then Eq.\,(\ref{eq:orderparam}) is the relevant one. In these conditions, using Eqs.\,(\ref{eq:orderparam}) and (\ref{eq:HBiomass}), one is able to determine the coefficient
\begin{equation}
	b = {1 \over 4} \alpha \Delta m_\text{LB} \langle \psi \rangle^{-3}.
	\label{bcoef}
\end{equation}
This provides enough information to obtain a predictive model
\begin{equation}
	\Delta F = b \psi^4 - 4 b \langle \psi \rangle^3 \psi,
\end{equation}
where $b$ is given by Eq.\,(\ref{bcoef}), allowing one to determine the free energy curve from Fig.\,\ref{fig:ESLGTH}. Taking the temperature shift measured in the last two centuries, of approximately $1~{\rm K}$ \citep{Jones:2004}, then $\psi \simeq \langle \psi \rangle = 1/T_{\rm H}$, leading to $b = (1 / 4) \alpha \Delta m_\text{LB} {T_{\rm H}}^3$.

The usefulness of the ES description based on a phase transition should not be limited to the transition from the Holocene to the Anthropocene. It can also be used, as shown in Fig.\,\ref{fig:ESLGT}, to set constraints on the ES for the negative minimum of the asymmetric phase with the last glaciation that occurred in the late Pleistocene. If one assumes this, then one can use the $q<0$ solution in Eq.\,(\ref{eq:minimawithoutH}), with the order parameter given by the temperature variation relative to the Holocene of about $5~{\rm K}$ \citep{Annan:2013}, to constrain the relation between the $a(q)$ and $b(q)$ coefficients to $a(q) = - 50 b(q)$.

During the Pleistocene, the ES was clearly controlled by natural processes. This shows how specific were the conditions of the Holocene, with $q = 0$ and how vulnerable the ES was to the unbalancing forces due to human activities. This assumption is consistent with an ES that features of a tipping point, as suggested by some of the existing data \citep{Lenton:2008,Lenton:2013}.

This example with a single component shows how the model here proposed can be used with data for the multiple components of the impact of human activities on the ES. Combined with more detailed data and the temperature record, the model can be constrained to allow for making numerical predictions. The natural evolution of this model is to become a multi-parameter description of the ES, following the path already being paved by attempts to identify the multiple components of the ES and human influence upon it, in models such as IPAT, ImPACT and SIRPAT \citep{York:2003}.

In Ref.\,\citep{Gaffney:2017}, human intervention is discriminated into components $P$, $C$ and $T$, population, consumption and technosphere, respectively; the latter are further divided into another three components, as already mentioned in the Introduction. In the proposed model, this would lead to equilibrium states driven, at lowest order, by a set of linear terms in $\psi$ in the free energy or by a single involved function of all relevant variables times $\psi$. However, since most of the components of $H$ are growing functions, there would be no change in the overall picture of the Anthropocene as a new state of the ES and the conclusions stated above would hold. Of course, only at the level of the Technosphere one can expect a counter balancing factor. However, the effect of the technosphere on the rate of change of the other human effects is rather modest and, in any case, is presumably effective only on a longer time scale. Until then, and according to the model presented here, the ES will remain evolving towards an unbalanced state in an accelerated evolution towards a new equilibrium state that is shifting further away from the Holocene equilibrium conditions.

The same conclusions would hold if the planetary boundaries picture were adopted. In this approach, $H$, would be a function of nine variables \citep{Steffen:2015}: rate of biosphere loss, land system change, global fresh water use, biogeochemical flows (global $N$ and $P$ cycles), ocean acidification, atmospheric aerosol loading, stratospheric ozone depletion, climate change, and chemical pollution. Hence, the Anthropocene would correspond all the same to a state where temperatures diverge more and more from the average Holocene one, provided $H$, as a function of the planetary boundaries variables, does not change sign. Thus, given that the first two derivatives of the above nine variables seem to remain positive, the predictions spelled out above will not change.

Furthermore, the quantitative insight provided by the LGT suggests that the Anthropocene is an inevitable new state in the history of Earth and humanity given the current socio-economic system and its underlying values. Only through a drastic change in this system will humankind be able to avert the menacing implications of the Anthropocene.

%-----------------------------------------------------------------------------%
%-- Bibliography -------------------------------------------------------------%
%-----------------------------------------------------------------------------%

\section*{References}

\bibliographystyle{elsarticle-harv}

\bibliography{anthropocene_landau}

\end{document}